\documentclass[prd,showpacs,showkeys,twocolumn,nofootinbib]{revtex4}
\usepackage{graphicx}

\newcommand{\beq}{\begin{equation}}
\newcommand{\eeq}{\end{equation}}
\newcommand{\bea}{\begin{eqnarray}}
\newcommand{\eea}{\end{eqnarray}}

\newcommand{\Ref}[1]{(\ref{#1})}
\newcommand{\erf}{\mathop{\rm erf}}

\begin{document}

\title{Wormholes supported by a phantom energy}

\author{Sergey Sushkov}%
\email{sergey_sushkov@mail.ru}%
\affiliation{Department of Mathematics, Kazan State Pedagogical
University, Mezhlauk 1 str., Kazan 420021, Russia}

\begin{abstract}
We extend the notion of phantom energy---which is generally accepted for
homogeneously distributed matter with $w<-1$ in the universe---on
inhomogeneous spherically symmetric spacetime configurations. A
spherically symmetric distribution of phantom energy is shown to be able
to support the existence of static wormholes. We find an exact solution
describing a static spherically symmetric wormhole with phantom energy and
show that a spatial distribution of the phantom energy is mainly
restricted by the vicinity of the wormhole's throat. The maximal size of
the spherical region, surrounding the throat and containing the most part
of the phantom energy, depends on the equation-of-state parameter $w$ and
cannot exceed some upper limit.
\\


\end{abstract}

\maketitle

\section{Introduction}
Recent astrophysical observations \cite{Rie-etal,Per-etal} related
to distant supernovas, cosmic microwave background, and galaxy
clustering all together essentially changed our view on the
evolution of the Universe. Now it is generally accepted that the
Universe at present is expanding with acceleration. The
explanation of such the `unexpected' cosmological behavior in the
framework of general relativity requires the supposition that a
considerable part ($\sim 70\%$) of the Universe consists of a
hypothetical {\em dark energy}: the exotic matter with a positive
energy density $\rho>0$ and a negative pressure $p=w\rho$ with
$w<-1/3$. In last few years intensive efforts have been undertaken
in modelling the dark energy (see the reviews
\cite{SahSta,Car,PeeRat,Sah:02,Pad,Sah:04}). A variety of
theoretical ideas and models concerning dark energy includes the
cosmological constant, quintessence, the Chaplygin gas, modified
gravity and scalar-tensor theories, braneworld models, dark energy
driven by quantum effects, k-essence, dark energy models with
negative potentials, tachyonic scalar fields, scalar fields with a
negative kinetic energy {\em etc}.

The most exotic form of dark energy is a {\em phantom energy} with
$w<-1$ \cite{Cal}, for which the weak energy condition is
violated. It is worth to note that values $w<-1$ not only are not
excluded but even are favored by recent observations
\cite{Ton-etal,Ala-etal,ChoPad,Alc}. The exotic nature of phantom
energy reveals itself in a number of unusual cosmological
consequences. One of them is a big rip \cite{CalKamWei}, i.e. a
final cosmological singularity to which the universe evolves
during a finite interval of time.
The related thermodynamical properties of a phantom universe are
also strange. Such the universe has a negative entropy diverging
near the big rip \cite{BreNojOdiVan} (an important role of quantum
effects near the big rip is discussed in \cite{NojOdi}), and a
negative temperature \cite{GonSig}. Another interesting phenomenon
is that all black holes in the phantom universe lose there masses
to vanish exactly in the big rip \cite{BabDokEro}.

If one takes seriously the phantom energy existence, one should expect
that its exotic nature would reveal itself not only on cosmological
scales. In particular, it is well known that the violation of the weak
energy condition is a necessary condition for existence of wormholes
\cite{MorTho,VisserBook}. Morris and Thorne in their seminal paper
\cite{MorTho} named the matter being able to support wormholes ``exotic''.
Therefore, one can consider the phantom energy as a possible candidate for
exotic matter. However, it is necessary to notice that if one try to
realize such the consideration in practice, one will face with a serious
problem. The point is that the generally accepted notion of dark/phantom
energy applies to an homogeneous distribution of matter in a universe.
Such the matter is characterized by the only two values: the energy
density $\rho$ and the negative homogeneous pressure $p$ related with each
other by the equation of state $p=w\rho$ with $w<-1/3$ (notice that the
equation-of-state parameter $w$ could generally speaking be variable). At
the same time, a wormhole spacetime is inhomogeneous, and so it demands a
non-homogeneously distributed matter. For example a spherically symmetric
wormhole needs a material characterizing by two different pressures:
radial and transverse. Thus, the question which should be answered is: Can
we extend the notion of dark/phantom energy on inhomogeneous spacetime
configurations?

In the recent paper \cite{SusKim} we discussed a model including a scalar
field with a negative kinetic term (a ghost scalar field), which is often
considered as a simple example of phantom energy \cite{SamTop}. In the
framework of the model we have obtained an exact time-dependent solution
describing a spherically symmetric wormhole in cosmological setting. The
wormhole was shown to connect two asymptotically homogeneous, spatially
flat universes expanding with acceleration. It is important that the
equation of state of the ghost scalar field can be effectively presented
as $p=w\rho$, where $p$ is a {\em radial} pressure, while a transverse
pressure $p_{tr}$ is only indirectly connected with the energy density via
field equations. Our analysis has revealed that the radial pressure is
everywhere and everywhen negative, and out of the wormhole's throat the
radial and transverse pressures tend quickly to be equal. This feature
indicates that out of the throat (i.e., in regions representing two
homogeneous spatially flat universes expanding with acceleration) the
ghost scalar field behaves effectively as dark energy providing the
accelerated expansion of the Universe.

The preceding analysis prompts us a way how to extend the notion of
phantom energy on the case of spherically symmetric spacetime
configurations. By analogy we will suppose that a spherically symmetric
distribution of phantom energy is characterized by the equation of state
$p=w\rho$ with $w<-1$, where $p$ is the negative radial pressure, while
the transverse pressure $p_{tr}$ is found from field equations. In this
paper we will use this approach to study static spherically symmetric
wormholes with phantom energy.


The paper is organized as follows. In the section \ref{II} we
briefly consider general properties of static spherically
symmetric wormholes. In the section \ref{III} we discuss a
spherically symmetric distribution of phantom energy and
demonstrate that it provides the flare-out conditions in the
wormhole's throat. An exact solution describing a static
spherically symmetric wormhole supported by the phantom energy is
constructed and analyzed in detail in the section \ref{IV}. The
section V summarizes the results obtained.

\section{Static spherically symmetric wormholes: Basic results}
\label{II}
The general metric of a static spherically symmetric Lorentzian wormhole
can be written down in Schwarzschild coordinates $(t,r,\theta,\varphi)$ as
follows \cite{MorTho,VisserBook}:
\beq\label{metric}
ds^2=-e^{2f(r)}dt^2+\frac{dr^2}{1-b(r)/r}+r^2
\left[d\theta^2+\sin^2\theta\,d\varphi^2\right].
\eeq
The properties of the wormhole geometry dictate some additional
requirements for the metric \Ref{metric}, which was in great detail
discussed in \cite{MorTho,VisserBook}. In particular, we note that
(i) the coordinate $r$ runs between $r_0\le r<+\infty$, where $r_0$ is the
throat radius. In order to cover the whole spacetime one have to use two
copies of the coordinate system \Ref{metric}.
(ii) The redshift function $f(r)$ must be everywhere finite; this
guarantees that no horizons exist in the spacetime.
(iii) The shape function $b(r)$ must obey the flare-out conditions at the
throat $r=r_0$:
\beq\label{beqr}
b(r_0)=r_0,
\eeq
and
\beq\label{flareout2}
b'(r_0)<1.
\eeq
(iv) Out of the throat, i.e., at $r>r_0$, $b(r)$ should satisfy
the following inequality:
\beq\label{blessr}
b(r)<r.
\eeq
(v) If one needs an asymptotical flatness of the spacetime
geometry one should require the limit
\beq\label{flatness}
b(r)/r\to 0\quad{\rm as}\quad |r|\to\infty.
\eeq

Because of the spherical symmetry the only nonzero components of the
stress-energy tensor are $T_0^0=-\rho(r)$, $T_1^1=p(r)$, and
$T_2^2=T_3^3=p_{tr}(r)$, where $\rho$ is the energy density, $p$ is the
radial pressure, and $p_{tr}$ is the transverse pressure. The Einstein
equations, $G_{\mu\nu}=8\pi T_{\mu\nu}$, now yield
%
\bea
\rho(r)&=&\frac{b'}{8\pi r^2};\label{eq1}\\
p(r)&=&\frac{1}{8\pi}\left[-\frac{b}{r^3}+2\frac{f'}{r}
\left(1-\frac{b}{r}\right)\right];\label{eq2}\\
p_{tr}(r)&=&\frac{1}{8\pi}\left(1-\frac{b}{r}\right)
\left[f''-\frac{b'r-b}{2r(r-b)}f'\right.\nonumber\\
&&\left.+f'^2+\frac{f'}{r} -\frac{b'r-b}{2r^2(r-b)}\right].\label{eq3}
\eea
%
Since the Einstein tensor obeys the identity $G^\alpha_{\beta;\alpha}=0$
and values of $\rho$, $p$ and $p_{tr}$ are connected by the conservation
law $T^\alpha_{\beta;\alpha}=0$:
\beq\label{conserv}
p'+f'\rho+\left(f'+\frac2r\right)p-\frac2r\,p_{tr}=0,
\eeq
the only two equations of the system (\ref{eq1}-\ref{eq3}) are
independent. It is convenient to represent them as follows:
\bea
b'&=&8\pi\rho\,r^2;\label{neweq1}\\
f'&=&\frac{8\pi p\,r^3+b}{2r(r-b)}.\label{neweq2}
\eea

The Einstein equations are connecting the geometrical flare-out conditions
\Ref{beqr}, \Ref{flareout2} with a distribution of matter in the wormhole
throat. In particular, supposing $b(r_0)=r_0$ in the equation \Ref{eq2} we
find
\beq\label{p0}
p_0\equiv p(r_0)=-\frac{1}{8\pi r_0^2}.
\eeq
Thus, the radial pressure at the throat should be negative to prevent it
from collapsing. Supposing $b'(r_0)<1$ in the equation \Ref{eq1} we obtain
\beq\label{rho0}
\rho_0\equiv\rho(r_0)<\frac{1}{8\pi r_0^2}.
\eeq
This inequality impose a restriction for the value of energy density at
the throat.

Out of the throat the equation \Ref{neweq1} is easily integrated:
\beq\label{b}
b(r)=r_0+\int_{r_0}^{r}8\pi\rho(\tilde r)\tilde r^2 d\tilde r.
\eeq
Here the constant of integration is chosen to provide the condition
$b(r_0)=r_0$. Instead of $b(r)$ one may consider the function
$m(r)=b(r)/2$ which is the effective mass inside the radius $r$. The limit
$\lim_{r\to\infty}m(r)=M$, if exists, represents the asymptotical wormhole
mass seen by an distant observer.

\section{Spherically symmetric distribution of phantom energy}
\label{III}
In addition to the Einstein equations one has also to specify an equation
of state for the matter being a source of gravity. The equation of state
describing phantom energy in cosmology is usually taken as $p=w\rho$ where
$w<-1$, and $p$ is a negative {\em spatially homogeneous} pressure. By
analogy we will suppose that a spherically symmetric distribution of
phantom energy is characterized by the equation of state in the same form,
$p=w\rho$, but now $p$ is the negative {\em radial} pressure, while the
transverse pressure $p_{tr}$ is defined by Eq. \Ref{conserv}. Denoting,
for convenience, $\kappa\equiv-w$ we have hereinafter
\beq\label{eqstate}
p=-\kappa\rho
\eeq
with $\kappa>1$.

An important feature of the spherically symmetric distributed
phantom energy is that it is able to provide the flare-out
conditions in the wormhole's throat. Really, if at the throat it
is fulfilled $p_0=-(8\pi r_0^2)^{-1}$, then
\beq\label{perho0}
\rho_0=-\frac{p_0}{\kappa}=\frac1{8\pi\kappa r_0^2}<\frac1{8\pi r_0^2},
\eeq
and hence both conditions \Ref{p0} and \Ref{rho0} are satisfied.

In the next section we will demonstrate that the phantom energy actually
can support wormholes and present two explicit solutions describing a
phantom energy wormhole.

\section{Phantom energy wormholes}
\label{IV}
The system of four equations \Ref{conserv}, \Ref{neweq1},
\Ref{neweq2}, and \Ref{eqstate} remains still to be incomplete
because we have five functions to be determined: $f$, $b$, $\rho$,
$p$, and $p_{tr}$. To solve this problem one must define one of
the functions ``by hand''. It seems reasonably to specify a
certain spatial distribution of the phantom energy density
$\rho(r)$. The form of $\rho(r)$ is only restricted by the
relation \Ref{eq1} and the conditions \Ref{flareout2},
\Ref{blessr}, \Ref{flatness}. Below we will analyze two examples
illustrating a various choice of $\rho(r)$.

\subsection{Phantom energy confined in a bounded spherical region}
First consider the simple model
\beq
\rho(r)=\left\{
\begin{array}{lc}
\rho_0,& r_0\le r\le r_1\\
0,& r>r_1
\end{array} \right.
\eeq
where $\rho_0$ is a constant. In this model the phantom energy is confined
in the bounded spherical region $r\le r_1$ (the {\em region I}) including
the wormhole's throat, while the region $r>r_1$ (the {\em region II}) is
supposed to be empty, so that $\rho=p=p_{tr}=0$ there.

Obtain a solution in the region I. For this aim we take into account that
$\rho_0=(8\pi\kappa r_0^2)^{-1}$ (see Eq. \Ref{perho0}). Substituting this
value of $\rho_0$ into \Ref{b} and integrating we can find the shape
function $b(r)$ in the region I in the following form:
\bea\label{bI}
b_I(r)&=&r_0+\frac{1}{3\kappa r_0^2}(r-r_0)(r^2+rr_0+r_0^2)\nonumber\\
&=&r-\frac{1}{3\kappa r_0^2}(r-r_0)(r-r_-)(r_+-r),
\eea
where
$$
r_\pm=\frac{r_0}{2}\left(\pm\sqrt{12\kappa-3}-1\right).
$$
Note that the inequality $b_I(r)<r$ would be satisfied for all
$r\in[r_0,r_1]$ only if
\beq\label{r1}
r_1<r_+=\frac{r_0}{2}\left(\sqrt{12\kappa-3}-1\right).
\eeq
It is worth to emphasize that the condition \Ref{r1} means that the region
containing the phantom energy {\em cannot} be arbitrarily large. Its
maximal size does not exceed $r_+$, which in turn depends on the
equation-of-state parameter $\kappa$.\footnote{This conclusion was made
for the radial coordinate $r$. One may check that the same is true for the
proper radial coordinate $l=\pm\int_{r_0}^{r}\frac{d\tilde
r}{\sqrt{1-b(\tilde r)/\tilde r}}$.}

Further, substituting the found expression for $b_I$ into \Ref{neweq2} we
can obtain the following expression for the redshift function $f(r)$:
\beq\label{fI}
e^{2f_I(r)}=C\Psi(r/r_0),
\eeq
where
\beq\label{Psi}
\Psi(x)=x^{-1}(x_+ -x)^{\frac{3\kappa x_-}{1+2x_-}}(x -
x_-)^{\frac{3\kappa x_+}{1+2x_+}},
\eeq
$C$ is a constant of integration, and $x_\pm=r_\pm/r_0$.

The solution in the region II has the Schwarzschild form:
\beq\label{bfII}
b_{{I\!I}}(r)=2M,\quad e^{2f_{I\!I}(r)}=1-\frac{2M}{r},
\eeq
where $M>0$ is a mass parameter.

\begin{figure}[t]\label{b1}
\centerline{\hbox{\includegraphics{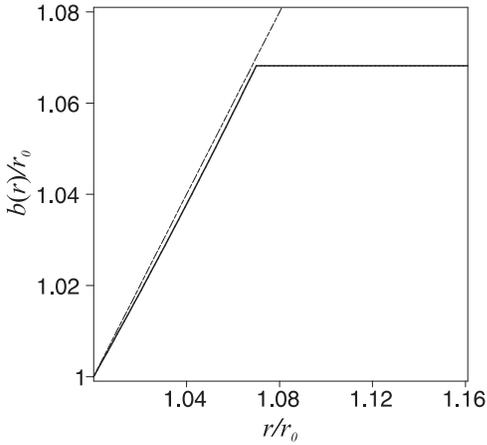}}} \caption{The diagram
represents the solution $b(r)/r_0$ given by the formulas \Ref{bI},
\Ref{bfII}, and \Ref{M} in case $\kappa=1.1$. The dashed straight line
corresponds to $r/r_0$.}
\end{figure}
\begin{figure}[t]\label{f1}
\centerline{\hbox{\includegraphics{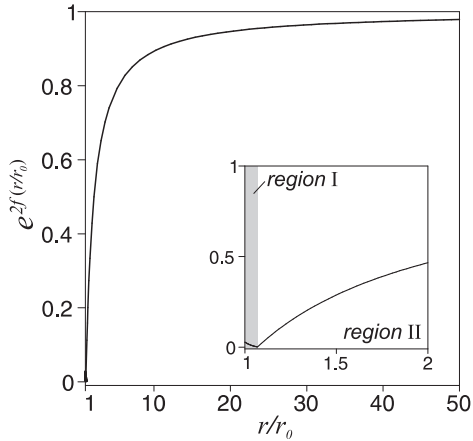}}} \caption{The graph of
$e^{2f(r/r_0)}$ is plotted for $\kappa=1.1$. The narrow shaded strip marks
the region I containing the phantom energy.}
\end{figure}
The formulas \Ref{bI}, \Ref{fI}, and \Ref{bfII} represent two separate
solutions for the regions I and II. To construct a solution in the whole
spacetime including both the region I and II we have to suppose the
continuity of the metric at the boundary $r=r_1$. Assuming
$\left.b_{I}(r)\right|_{r_1}=\left.b_{I\!I}(r)\right|_{r_1}$ yields
\bea\label{M}
2M&=&r_0+\frac{1}{3\kappa r_0^2}(r_1-r_0)(r_1^2+r_1r_0+r_0^2)\nonumber\\
&=&r_1-\frac{1}{3\kappa r_0^2}(r_1-r_0)(r_1-r_-)(r_+-r_1).
\eea
This relation expresses the wormhole mass $M$ via the throat's radius
$r_0$ and the size of region I, $r_1$. The second condition
$\left.f_{I}(r)\right|_{r_1}=\left.f_{I\!I}(r)\right|_{r_1}$ is fixing the
value of $C$ as follows
\beq\label{C}
C=\left(1-\frac{2M}{r_1}\right)\Psi^{-1}(r_1/r_0).
\eeq

Now, the functions $b_{I,I\!I}$ and $f_{I,I\!I}$ given by Eqs. \Ref{bI},
\Ref{fI}, \Ref{bfII} together with the matching conditions \Ref{M},
\Ref{C} form a solution describing a static spherically symmetric wormhole
supported by the phantom energy. In the figures 1 and 2 we give the
graphical representation of the obtained solution.

\subsection{The smooth phantom energy density distribution}
Now let us discuss a smooth distribution of the phantom energy density.
For this aim we will model the function $\rho(r)$ using the normal
Gaussian distribution law:
\beq\label{smoothrho}
\rho(r)=\rho_0 e^{-\alpha(r/r_0-1)^2},
\eeq
where $\rho_0=(8\pi\kappa r_0^2)^{-1}$ is the value of phantom energy
density at the throat, and $\alpha>0$ is a model parameter.

\begin{figure}[t]\label{b2}
\centerline{\hbox{\includegraphics{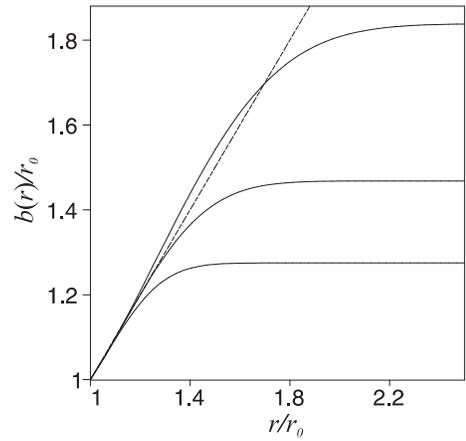}}} \caption{The diagram
represents the solution $b(r)/r_0$ (solid curves) corresponding to the
smooth phantom energy density distribution $\rho(r)=\rho_0
e^{-\alpha(r/r_0-1)^2}$ and given by the formula \Ref{smoothb} in case
$\kappa=1.1$ and $\alpha=3$, 6.68, 15 from top to bottom, respectively.
The corresponding critical value of $\alpha$ is $\alpha_*\approx 6.68$.
The dashed straight line represents $r/r_0$.}
\end{figure}
\begin{figure}[t]\label{f2}
\centerline{\hbox{\includegraphics{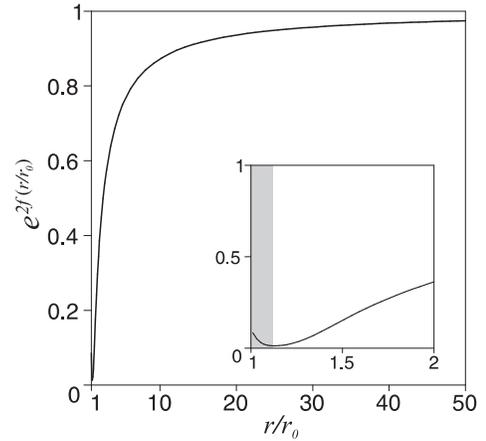}}} \caption{The graph of
$e^{2f(r/r_0)}$ is plotted for $\kappa=1.1$ and $\alpha=15$.}
\end{figure}

Substituting \Ref{smoothrho} into \Ref{b} and integrating yields
\bea
\frac{b(r)}{r_0}&=&1+\frac1{\kappa\alpha}-(x+1)\,e^{-\alpha(x-1)^2}\nonumber\\
&&+\frac{\sqrt{\pi}}{2\kappa\sqrt{\alpha}}\left[1+\frac{1}{2\kappa\alpha
 }\right]\erf(\sqrt{\alpha}(x-1)), \label{smoothb}
\eea
where $x\equiv r/r_0$, and $\erf(z) = 2\pi^{-1/2}\int_0^z e^{-t^2}dt$ is
the error function. The function $b(r)$ given by \Ref{smoothb}
automatically obeys the flare-out conditions \Ref{beqr}, \Ref{flareout2}
at the wormhole's throat. The additional condition \Ref{blessr},
restricting a behavior of $b(r)$ out of the throat, imposes a constraint
on the parameter $\alpha$. A typical behavior of $b(r)$ is illustrated in
the figure~3. One may see that a curve $b(r)/r_0$ is entirely situated
under the straight line $r/r_0$ in case $\alpha>\alpha_*$, where
$\alpha_*$ is some critical value of the parameter $\alpha$. That is the
condition $b(r)<r$ is satisfied for all $r>r_0$, i.e. everywhere out of
the throat, if and only if $\alpha>\alpha_*$. This imposes a certain
restriction on the spatial distribution of the phantom energy density.
Really, the value of $\alpha$ in \Ref{smoothrho} determines how fast
$\rho(r)$ is decreasing. Namely, $\rho(r)$ becomes $e$ times less than
$\rho_0$ at $r_1=r_0(1+\alpha^{-1/2})$, and then, for $r>r_1$, the value
of $\rho(r)$ is rapidly decreasing. In this sense one can regard the
region $r_0\le r\le r_1$ as that which concentrates the most part of the
phantom energy density. The size of this region is proportional to
$\alpha^{-1/2}$ and {\em cannot} exceed some maximal size
$\sim\alpha_*^{-1/2}$. (Note that, as follows from numerical analysis,
$\alpha_*$ depends ultimately on $\kappa$.) The asymptotical wormhole mass
$M$ reads
\bea
M&=&\lim_{r\to\infty}\frac12 b(r)\nonumber\\
&=& \frac12{r_0}\left\{1+\frac1{\kappa\alpha}
+\frac{\sqrt{\pi}}{2\kappa\sqrt{\alpha}}\left[1+\frac{1}{2\kappa\alpha
 }\right]\right\}.
\eea
It is worth to notice that $M$ is positive.

The redshift function $f(r)$ can be found numerically by using Eq.
\Ref{neweq2}. The figure~4 illustrates a typical behavior of $f(r)$.

\section{Concluding remarks}
\label{V}
In this paper we have constructed exact solutions describing a
static spherically symmetric wormhole supported by the phantom
energy and, thus, explicitly demonstrated that the phantom energy
can support the existence of static wormholes. The obtained
solutions have revealed an interesting and important feature of
the phantom energy wormholes. It turns out that a spatial
distribution of the phantom energy is mainly restricted by the
vicinity of the wormhole's throat; and the maximal size of the
spherical region, surrounding the throat and containing the most
part of the phantom energy, cannot exceed some upper limit
depending on the equation-of-state parameter $\kappa$. Thus, the
phantom energy looks like to be confined near the wormhole's
throat. In this connection we notice that, since the asymptotical
mass $M$ of the phantom energy wormhole is positive, a distant
observer could not see a difference (of gravitational nature)
between such the wormhole and a compact mass $M$.

\section{Acknowledgement}
This work was supported by the Russian Foundation for Basic Research grant
No 05-02-17344.

\end{document}